\documentclass[epj]{svjour}
\usepackage{epsf}
\usepackage{amssymb}
\begin{document}
\title{
Finite-size-scaling analyses of the
chiral order in the Josephson-junction ladder
with half a flux quantum per plaquette
}
\titlerunning{
Chiral order in the Josephson-junction ladder
}
\author{Yoshihiro Nishiyama}
\institute{
Department of Physics, Faculty of Science,
Okayama University,
Okayama 700-8530, Japan}
\date{Received: date / Revised version: date}
%
\abstract{
Chiral order of
the Josephson-junction ladder with half a flux quantum per plaquette is
studied by means of the exact diagonalization method.
We consider an extreme quantum limit where each superconductor
grain (order parameter) is represented by $S=1/2$ spin.
So far,
the semi-classical $S\to\infty$ case, where each spin reduces
to a plane rotator, has been considered extensively.
We found that in the case of $S=1/2$,
owing to the strong quantum fluctuations,
the chiral (vortex lattice) order becomes dissolved
except in a region, where 
attractive intrachain and, to our surprise,
repulsive interchain interactions both exist.
On the contrary, for considerably wide range of parameters,
the superconductor ($XY$) order is kept critical.
The present results are regarded as a demonstration of the critical phase
accompanying chiral-symmetry breaking
predicted for frustrated $XXZ$ chain field-theoretically.
} 
%
\maketitle
\section{Introduction}
\label{section_Introduction}
By means of a field-theoretical technique,
Nersesyan, Gogolin and E{\ss}ler claimed 
\cite{Nersesyan98} that
the ground state of
the $XXZ$ spin chain with sufficiently strong 
next-nearest-neighbor interaction
is in the chiral phase, where
the spin-screw chirality is
broken spontaneously. 
Their proposal is astonishing, because 
the $XXX$ counterpart, that has been studied very extensively so far
\cite{Haldane82,Tonegawa87,Harada88,White96,Allen97,Harada99},
is known to be in the dimer phase,
where the translational symmetry is broken spontaneously
so long as the frustration is strong enough.
New treatments devised specifically for $XXZ$ were reported
\cite{Cabra00,Allen99}
so as to support the aforementioned prediction.   
Meanwhile,
as for other frustrated $XXZ$ chain, namely, the
three-leg ladder with diagonal interchain interaction,
Azaria {\it et al.}  argued \cite{Azaria98,Azaria99} that
a unique critical phase possessing high central charge ($c=2$) might 
be realized in the ground state.
According to their argument,
the chiral-symmetry breaking is significant to stabilize such
high-$c$ criticality.
These recent developments tell that the $XXZ$ anisotropy
with in-plane frustration may 
give rise to unexpected exotic phases.
The above-mentioned proposals are all based on field-theoretical
descriptions.
Therefore, in order to support these scenarios,
numerical simulations should be carried out.
Kaburagi {\it et al.} 
employed the exact diagonalization
and the density-matrix renormalization-group methods
for the $XX$ model with the next-nearest-neighbor coupling
\cite{Kaburagi99,Hikihara00}.
They concluded that the chirality is {\em not} broken for the $S=1/2$ chain,
whereas it would be broken for the $S=1$ chain;
namely, numerical-simulation result for $S=1/2$ appears to contradict
 the field-theoretical description.
Nevertheless, numerical simulation of such frustrated chain 
itself is a matter of serious methodological concern
\cite{White96};
main obstacles are due to the exponentially small energy gap and 
incommensurability of the spin-correlation function, 
which often miss-matches
the total system sizes.

In order to avoid such complications that would arise from
incommensurability,
we have investigated the two-leg Josephson-junction ladder,
\begin{eqnarray}
{\cal H} &=&
    -\frac{t_\parallel}{2}
                   \sum_{i=1}^L ({\rm e}^{{\rm i}\Phi} a^\dagger_{1i}a_{1,i+1}
           +{\rm e}^{-{\rm i}\Phi} a^\dagger_{1,i+1}a_{1i})      \nonumber \\
 & &-\frac{t_\parallel}{2}
                   \sum_{i=1}^L ( a^\dagger_{2i}a_{2,i+1}
                                 + a^\dagger_{2,i+1}a_{2i})      \nonumber \\
 & & -\frac{t_\perp}{2}
                   \sum_{i=1}^L ( a^\dagger_{1i}a_{2i}
                                 + a^\dagger_{2i}a_{1i})         \nonumber \\
\label{Hamiltonian}
 & & -V_\parallel \sum_{l=1,2} \sum_{i=1}^L
                 a^\dagger_{li}a_{li}a^\dagger_{l,i+1}a_{l,i+1}  \nonumber \\
 & & -V_\perp \sum_{i=1}^L 
                 a^\dagger_{1i}a_{1i}a^\dagger_{2i}a_{2i}   ,
\end{eqnarray}
with gauge-twist angle,
\begin{equation}
\Phi=\pi .
\end{equation}
Here, $a_{li}$ and $a^\dagger_{li}$
are the hard-core boson annihilation and creation operators, respectively,
at the $i$-th site along the $l$-th leg.
The parameters $t_\parallel$, $t_\perp$, $V_\parallel$ and $V_\perp$
control the intrachain hopping amplitude, the interchain hopping amplitude,
the nearest-neighbor intrachain attraction and the 
interchain attraction, respectively.
The number of bosons is set to $N=L$ (half filled).
Throughout this paper, we choose $t_\parallel$ 
as the unit of energy; namely,
\begin{equation}
t_\parallel=1.
\end{equation}
We have imposed the periodic-boundary condition along the ladder;
$a_{l,L+1}=a_{l,1}$.
The angle $\Phi$ denotes the gauge twist around each plaquette.
Because the angle is set to $\Phi=\pi$, 
the ladder is subjected to a uniform magnetic field of half
a flux quantum per plaquette.
That is, suppose that the bosons are in the superconducting state
(this is a subtle issue in one dimension),
the magnetic flux would possibly be quantized 
so as to form rigid vortex-lattice order along the ladder.
This order is confirmed to develop 
in the preceding semi-classical
analysis \cite{Ciria99}.
That is, the chiral order of our model has
two-unit-cell periodicity.
For the purpose of studying the stability of
chirality against the quantum fluctuation numerically,
our model 
is far more advantageous than the frustrated spin chain,
because
the spin-screw chiral order of the latter model
has long-wavelength incommensurate
structure, and thereby
the numerical data suffer from
insystematic finite-size-scaling behavior.
Tendency to the formation of such vortex structure may become more transparent,
if we transform the boson Hamiltonian (\ref{Hamiltonian})
in terms of spin:
Through utilizing the mapping relations between the hard-core boson and
the spin-$1/2$ operators, namely,
$a_{li}=S^-_{li}$, $a^\dagger_{li}=S^+_{li}$ and 
$a^\dagger_{li}a_{li}-1/2=S^z_{li}$,
the above boson model is mapped to the $XXZ$ spin ladder,
\begin{eqnarray}
{\cal H} &=& t_\parallel \sum_{i=1}^L (S^x_{1i}S^x_{1,i+1}+S^y_{1i}S^y_{1,i+1})
                                               \nonumber \\
  & & -t_\parallel \sum_{i=1}^L (S^x_{2i}S^x_{2,i+1}+S^y_{2i}S^y_{2,i+1})
                                               \nonumber \\
  & & -t_\perp \sum_{i=1}^L (S^x_{1i}S^x_{2i}+S^y_{1i}S^y_{2i}) \nonumber \\
\label{Hamiltonian_spin}
  & & -V_\parallel \sum_{l=1,2} \sum_{i=1}^L S^z_{li}S^z_{l,i+1}
  -V_\perp \sum_{i=1}^L S^z_{1i}S^z_{2i}  ,
\end{eqnarray}
apart from a constant term.
Note that the signs of the 
$XY$-component magnetic interactions
are different from one leg
and the other, and thus an in-plane frustration does exist.
Therefore, vortex-antivortex alignment gets favored.
The semi-classical version $S\to\infty$ of the model (\ref{Hamiltonian_spin})
has been studied extensively
\cite{Granato92,Granato93,Denniston95,Granato96,Ciria99}:
In the limit, spins are represented by rotators, and through
path-integral mapping, the system
reduces to a two-dimensional
classical rotator model.
According to numerical-simulation studies 
\cite{Granato92,Granato93,Granato96},
that rotator model appears to be in the KT critical phase.
To our surprise, surrounded by such critical background of in-plane spin components,
rigid long-range chiral order develops. 
And so, it is our motivation to investigate the stability of the chiral order
against the quantum fluctuations of $S=1/2$.
We show in this paper, that
the chiral order fails to develop except in a limited condition,
where an attractive intrachain coupling and a {\em repulsive} interchain
coupling are both turned on.
This new phase is a demonstration 
of critical phase accompanying chiral-symmetry
breaking predicted field-theoretically
\cite{Nersesyan98,Cabra00,Allen99}.

Let us mention some remarks concerning the present ladder model 
described by either boson representation eq. (\ref{Hamiltonian}) 
or spin representation eq. (\ref{Hamiltonian_spin}).
One might wonder that the hard-core condition,
in other words, the $S=1/2$ condition, would be too restrictive,
and the model misses microscopic physical ingredients such as 
the intra-and-inter-grain charge capacitances.
Actually, setup of our model might be rather phenomenological.
We stress, however, that the
one-dimensional $XXZ$ chains
belong to the Tomonaga-Luttinger-liquid
universality class irrespective of $S$ \cite{Alcaraz92}.
(Heuristically, it is known that one-dimensional quantum systems
possessing the ${\rm U}(1)$ symmetry are flown to that universality in many cases.)
As a matter of fact, according to the proposal \cite{Alcaraz92},
the transverse-correlation exponent $\eta_\perp$ is governed by the
compact formula,
\begin{equation}
\label{Alcaraz_Moreo}
\eta_\perp = \frac{\pi - \cos^{-1}J^z}{2 \pi S}           .
\end{equation}
(In our notation, $J^z$ is given by $J^z=-V_\parallel$.)
That is, through varying $J^z$,
one can cover all possible range of $\eta_\perp$
with $S=1/2$ {\em fixed}.
Lastly, we mention the possibility of the so-called
`spin liquid' state that often arises in quantum spin systems;
here, the term `spin liquid' denotes the state
with exponentially decaying
short-range correlation function.
we use the term spin liquid that
The conventional two-leg ladder 
\cite{Dagotto92,Dagotto96}
(with non-frustrated interchain coupling)
is a prototypical system exhibiting spin-liquid state.
We will show that our frustrated ladder does exhibit spin-liquid phase
for a certain parameter condition beside the chiral phase.

This paper is organized as follows.
In the next section, we explore the ladder model by
means of the exact diagonalization method.
The last section is devoted to summary and discussions.

\section{Numerical results}
\label{section_results}
In this section, we present numerical results.
We carried out exact-diagonalization calculations
for the system (\ref{Hamiltonian_spin}) with up to $N=2L=32$ spins.
The data are analyzed in terms of the finite-size-scaling 
theory.

\begin{figure}[htbp]
\begin{center}\leavevmode
\epsfxsize=8.5cm
\epsfbox{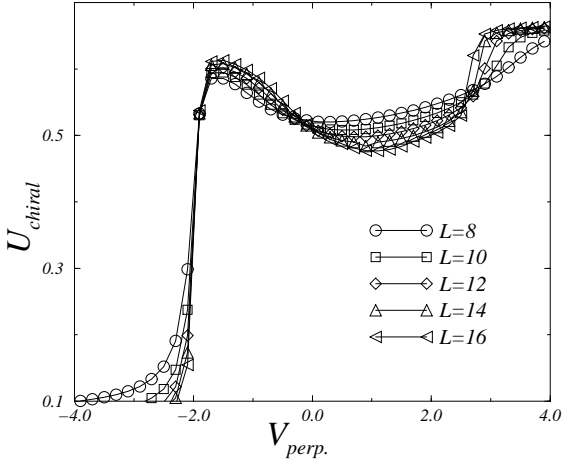}
\end{center}
\caption{
Binder parameter for the chiral order
({\protect \ref{Binder_chiral}}) is plotted for 
$t_\parallel=1$, $t_\perp=0.5$, $V_\parallel=0.6$
and various $V_\perp$.
The symbols $\circ$, $\square$, $\diamond$, $\triangle$
and $\triangleleft$ denote the data for the system sizes
$L=8$, $10$, $12$, $14$ and $16$, respectively. 
We see that
the chiral order develops in the repulsive
interchain coupling $-1.8 \lesssim V_\perp \lesssim 0$.
}
\label{binderCH_Jzp}
\end{figure}

In Fig. \ref{binderCH_Jzp}, we have plotted the Binder parameter
\cite{Binder81a,Binder81b},
\begin{equation}
\label{Binder_parameter}
U=1-
   \frac{\langle M^4 \rangle}
      {3 \langle M^2 \rangle^2}  ,
\end{equation}
for the chiral order,
\begin{eqnarray}
M=M_{\rm chiral} &=& 
              \sum_{i=1}^{L} (-1)^i 
   \left[ {\bf S}_{1i} \times {\bf S}_{2i} \right]_z   \nonumber \\
           &=& \sum_{i=1}^{L} (-1)^i (S^x_{1i}S^y_{2i}-S^y_{1i}S^x_{2i})   \nonumber \\
\label{Binder_chiral}
           &=& \frac{1}{2{\rm i}} \sum_{i=1}^{L} (-1)^i (-a_{1i}^\dagger a_{2i}
                                                  +a_{1i} a_{2i}^\dagger)  ,
\end{eqnarray}
for the system (\ref{Hamiltonian}) with 
$t_\perp=0.5$, $V_\parallel=0.6$ and $V_\perp$ varied.
Here, $\langle \cdots \rangle$ denotes the ground-state average.
Note that in the boson language, $M_{\rm chiral}$ measures
the staggered boson current through the rungs.
Therefore, it detects the formation of the vortex-lattice order.
Finite-size scaling behavior of the Binder parameter contains
the information
whether the order $M$ develops or not \cite{Binder81a,Binder81b}:
If the order is long (short) ranged, the
Binder parameter grows (becomes suppressed)
through enlarging the system sizes.
At critical point, the Binder parameter remains scale-invariant.
From Fig. \ref{binderCH_Jzp},
we found that the chiral order develops for the parameter 
range $-1.8 \lesssim V_\perp \lesssim 0$, while it is short ranged otherwise.
As would become more apparent in the succeeding analyses,
the strong-coupling regions of $V_\perp \gtrsim 2.5$ and $V_\perp \lesssim -1.8$
belong to
insulator phases of different characters,
and are thus rather out of present concern.
For $V_\perp \gtrsim 2.5$, for instance, the bosons are so cohesive that
they may constitute an island in the sea of vacant sites;
namely, the system becomes `phase separated.'
On the other hand, for $V_\perp \lesssim -1.8$, owing to the strong repulsion,
a Mott gap opens between the bonding and anti-bonding 
excitation branches.

Our result shows that
the region of the chiral phase is  limited.
Hence,
we found that
quantum fluctuations are dominating so
that
unlike the semi-classical case,
assistance of the many-body correlations of $V_\parallel$ and $V_\perp$ 
is
vital for stabilizing the chiral order.
After scanning the parameter space,
we found that the optimal condition lies around
$t_\perp=0.5$, $V_\parallel=0.6$ and $V_\perp=-1$.
It is fairly reasonable that the optimal intrachain interaction
is attractive, because attractive interaction enhances the
tendency toward superconductivity ($XY$ order); 
see eq. (\ref{Alcaraz_Moreo}), for instance.
On the contrary, the fact that the {\em repulsive} interchain
interaction is favorable sounds astonishing.
It is expected that the repulsive interchain coupling
might enhance the particle exchange between the chains.
Hence, we see that such particle exchange is significant in order
to confine (pin) the magnetic fluxes at each plaquette.

In order to confirm the above phase diagram,
we have calculated the scaled domain-wall energy,
\begin{equation}
\label{domain_wall_energy}
L \Delta E (L)=L 
\left(
E_{\rm g}(L)-\frac{L E_{\rm g}(L+1)}{L+1}
\right) .
\end{equation} 
\begin{figure}[htbp]
\begin{center}\leavevmode
\epsfxsize=8.5cm
\epsfbox{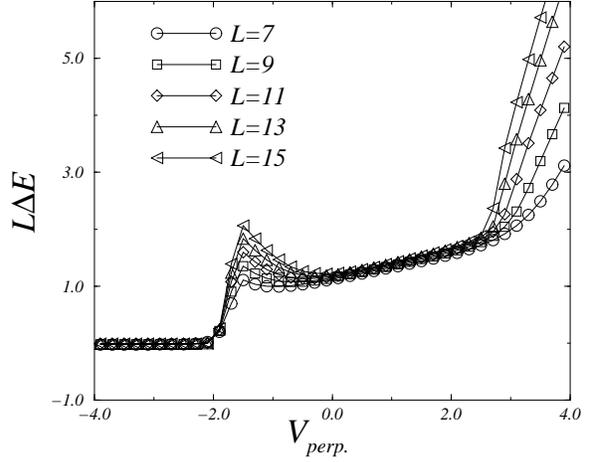}
\end{center}
\caption{
Scaled kink energy 
({\protect \ref{domain_wall_energy}}) is plotted
for the same parameter range as that of Fig.
{\protect \ref{binderCH_Jzp}}.
The symbols $\circ$, $\square$, $\diamond$, $\triangle$
and $\triangleleft$ denote the data for the system sizes
$L=7$, $9$, $11$, $13$ and $15$, respectively. 
In the region $-1.8 \lesssim V_\perp \lesssim 0$, 
where the chiral phase is realized (see Fig. 
{\protect \ref{binderCH_Jzp}}),
the scaled kink energy grows actually.
On the contrary, in the attractive region $V_\perp \gtrsim 0$,
the scaled kink energy stays scale-invariant.
This indicates that
a critical phase, probably of the $XY$ mode, is realized;
see Fig. {\protect \ref{binderXY_Jzp}}.
}
\label{jiheki_Jzp}
\end{figure}
$E_{\rm g}(L)$ denotes the ground-state energy of the system with size $L$,
which is supposed to be odd integer.
For the system with odd number of plaquettes,
one domain wall is created in the alternating
alignment of vortex-antivortex structure.
As would be apparent from the definition (\ref{domain_wall_energy}),
$\Delta E$ measures the extra domain-wall
energy cost due to the kink.
Hence, for ordered phase, the scaled kink energy should increase linearly with
respect to $L$, while for disordered phase, $L\Delta E$ vanishes exponentially;
namely,
$\sim {\rm e}^{-L/\xi}$ with correlation length $\xi$.
At critical point, $L \Delta E$ should be scale-invariant, 
because $\Delta E$ is of scaling dimension $1/L$; 
note the relations $\Delta E \sim 1/\xi \sim 1/L$.
In Fig. \ref{jiheki_Jzp},
we plotted $L \Delta E$ for the same parameter range
as that of Fig. \ref{binderCH_Jzp}.
In fact, in the chiral phase 
$-1.8 \lesssim V_\perp \lesssim 0$, 
which is estimated from the above Binder-parameter
analysis,
we observe clear signature of the
chiral-domain-wall energy cost.
On the other hand, in the area $0 \lesssim V_\perp \lesssim 2.5$, 
$L\Delta E$ seems to be scale-invariant.
In the former analysis of Fig. \ref{binderCH_Jzp},
we have concluded that in the region,
the chirality is disordered.
Hence, we conclude that in $0 \lesssim V_\perp \lesssim 2.5$,
the $XY$ order, rather than the chiral order,
exhibits criticality.
As are mentioned above, 
the strong-coupling regions of $V_\perp \gtrsim 2.5$ and $V_\perp \lesssim -1.8$
are belonging to insulator phases with different characters.
In terms of the pictures presented above,
the behaviors of $L\Delta E$ in these regions are readily understandable.
The blowup of $L \Delta E$ for $V_\perp \gtrsim 2.5$
is due to a kink (dislocation) formed in the island of particles;
note that the particle occupation number is forced
to be half-filled.
On the other hand,
the rapid closure of $L\Delta E$ for $V_\perp \lesssim -1.8$
is precisely due to the fact that the system is a Mott insulator.

\begin{figure}[htbp]
\begin{center}\leavevmode
\epsfxsize=8.5cm
\epsfbox{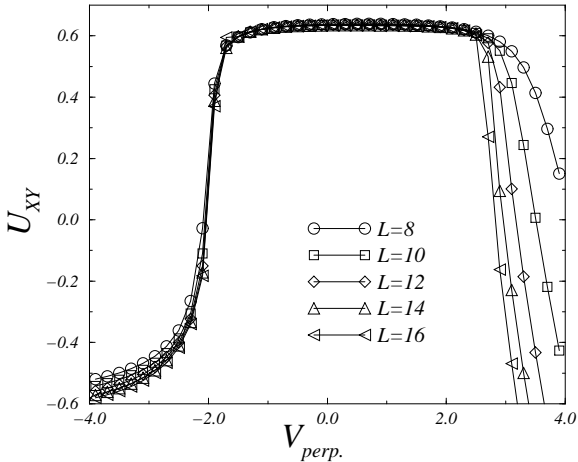}
\end{center}
\caption{
Binder parameter for the $XY$ order
({\protect \ref{Binder_XY}}) is plotted for 
$t_\parallel=1$, $t_\perp=0.5$, $V_\parallel=0.6$
and various $V_\perp$.
The symbols $\circ$, $\square$, $\diamond$, $\triangle$
and $\triangleleft$ denote the data for the system sizes
$L=8$, $10$, $12$, $14$ and $16$, respectively. 
The $XY$ order appears to be critical in the region
$-1.8 \lesssim V_\perp \lesssim 2.5$.
In the region, the fixed point value of
the Binder parameter seems to be unchanged.
Hence,
we see that the $XY$ order is not influenced 
by the change of $V_\perp$.
}
\label{binderXY_Jzp}
\end{figure}

Let us turn to the $XY$ order.
In Fig. \ref{binderXY_Jzp}, we plotted the Binder parameter 
(\ref{Binder_parameter}) for the in-plane
spontaneous magnetization ($XY$ order),
which corresponds to the 
superconductivity 
(gauge coherence)
order parameter
in the boson language,
\begin{equation}
\label{Binder_XY}
M^2=M_{XY}^2=\sum_{limj} (S^x_{li}S^x_{mj}+S^y_{li}S^y_{mj}) ,
\end{equation}
for the same parameter range as that of former figures.
From the plot, we observe that the $XY$ order is kept critical
for considerably wide range of parameter 
$-1.8 \lesssim V_\perp \lesssim 2.5$.
We notice that
this $XY$ critical phase does contain the chiral phase.
This feature is quite contrastive to that of the semi-classical case
$S\to\infty$, where the chirality is stronger (more stable) than
the $XY$ order, and thus the chiral phase contains the $XY$ phase
\cite{Granato93}.
Moreover, it should be noted that in the critical region, the fixed-point
value of the Binder parameter is kept hardly changed.
In consequence,
we found that the $XY$ correlation function is not influenced
very much by $V_\perp$,
while
the chiral sector is affected significantly by $V_\perp$.
These results show that those two sectors behave independently;
in the commonly referred terminology,
those two sectors are `separated.'
Such the situation where each mode is described by respective
low-energy effective theory occurs commonly in one-dimensional physics.
This point is discussed in Section \ref{section_summary}.

\begin{figure}[htbp]
\begin{center}\leavevmode
\epsfxsize=8.5cm
\epsfbox{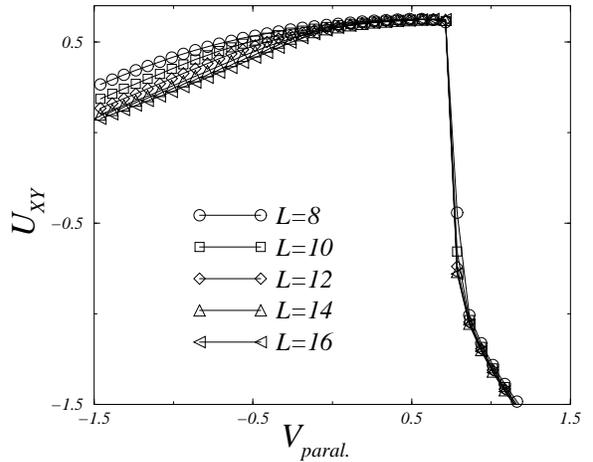}
\end{center}
\caption{
Binder parameter for the $XY$ order
({\protect \ref{Binder_XY}}) is plotted for 
$t_\parallel=1$, $t_\perp=0.5$, $V_\perp=-1$
and various $V_\parallel$.
The symbols $\circ$, $\square$, $\diamond$, $\triangle$
and $\triangleleft$ denote the data for the system sizes
$L=8$, $10$, $12$, $14$ and $16$, respectively. 
The $XY$ order becomes disordered eventually in the repulsive
intrachain coupling $V_\parallel \lesssim 0$
and driven to spin-liquid phase.
}
\label{binderXY_Jzh}
\end{figure}

Lastly, let us turn our attention to the possibility of
spin-liquid state. 
The simplest way to realize the spin-liquid state is given
just by setting
$t_\perp$ to a very large value.
In that strong-interchain-coupling limit, the bonding and
anti-bonding excitation branches
are separated so that a band gap opens and thereby the ground state
becomes disordered.
Other than that rather trivial way,
we found that the spin-liquid state is accessible
from the set of parameters treated in the former figures
\ref{binderCH_Jzp}-\ref{binderXY_Jzp}
just through setting $V_\parallel$ repulsive;
see Fig. \ref{binderXY_Jzh}, 
where we plotted the Binder parameter for the $XY$ order
for $t_\perp=0.5$, $V_\perp=-1$ and various $V_\parallel$.
We see that for attractive coupling $0 \lesssim V_\parallel \lesssim 0.7$, 
the $XY$ sector stays critical as
is presented before.
(The rapid suppression of $U_{XY}$ in $V_\parallel \gtrsim 0.7$ 
is due to phase separation caused by strong attraction.)
On the other hand, for repulsive coupling $V_\parallel \lesssim 0$,
the $XY$ order
becomes disordered eventually and driven to a spin-liquid phase.
This behavior tells that the interchain coupling
is a relevant perturbation in 
$V_\parallel \lesssim 0$.
That feature coincides with that of the conventional non-frustrated ladder
\cite{Schulz86,Strong92,Strong94}.
We discuss this similarity in the next section.

\section{Summary and discussions}
\label{section_summary}
We have diagonalized the Josephson-junction ladder
subjected to the uniform magnetic field of half a flux quantum per plaquette
(\ref{Hamiltonian}).
Unlike the semi-classical case, the chiral order suffers 
from strong disturbances
due to quantum fluctuations.
In order to stabilize the chiral order, we need to tune carefully
the many-body interactions
such as 
$V_\parallel=0.6$ and $V_\perp=-1$.
It is surprising that the {\em repulsive} interchain interaction
gives rise to the stabilization of the chiral order.
This fact indicates that the particle exchange 
across the chains over the rungs,
rather than the gauge coherence,
is vital for pinning the vortices at each plaquette.
On the contrary, we found that the $XY$ order 
(\ref{Binder_XY}) is insensitive to 
$V_\perp$;
the $XY$ order is kept critical for considerably wide range of parameters.
This result indicates that the chiral and $XY$ sectors are
separated.

In quantum spin systems, owing to the strong quantum fluctuations,
the so-called spin-liquid state can appear \cite{Dagotto92}.
The spin-liquid state, in fact, 
has been the main concern 
in the course of studies of the (non-frustrated) two-leg ladder
\cite{Dagotto96}.
Close to the chiral phase mentioned above,
we found that a spin-liquid state emerges just through 
changing the sign of $V_\parallel$.
Surprisingly,
this behavior coincides with that of the conventional ladder;
with the bosonization method 
\cite{Schulz86,Strong92,Strong94},
it was shown that the interchain
coupling becomes relevant for $V_\parallel \lesssim 0$ 
(antiferromagnetic intrachain interaction).
Hence,
it is suggested that 
the $XY$ mode of our frustrated ladder and 
that of the conventional ladder
behave similarly.
This similarity might be reasonable, if we 
remember the separation of the chiral and $XY$ sectors mentioned above.
Hence, it is suggested that the present critical phase,
as well as that of conventional ladder 
\cite{Schulz86,Strong92,Strong94},
belongs to the universality class
of the central charge $c=1$.
\cite{Schulz86,Strong92,Strong94}.
Direct identification of the universality class
in terms of the finite-size conformal-field theory, for instance,
may be exceedingly
troublesome,
because the excitation structure of the $XY$ sector is smeared by
that of the chiral sector.
This would remain for future study.

\section*{Acknowledgments}
The author is grateful to Professor I. Harada
for helpful discussions.

\end{document}